\documentclass[journal,12pt,draftclsnofoot,onecolumn]{IEEEtran}

\usepackage{cite}
\usepackage[dvips]{graphicx}
\usepackage{multicol}

\usepackage{amsfonts}
\usepackage{amssymb}
\usepackage{amsmath}
\usepackage{floatflt}

\begin{document}
\title{On the response of an antenna to polarized electromagnetic plane waves using a tensorial and spinorial approach}
\author{David~Bebbington~\IEEEmembership{} and~Laura~Carrea~\IEEEmembership{}
\thanks{
        This work was supported by the Marie Curie Research Training
Network AMPER (Contract number HPRN-CT-2002-00205).}
\thanks{D. Bebbington and L. Carrea are with the Centre for Remote Sensing \& Environmetrics, School of Computer Science and Electronic Engineering, University of Essex, U.K. (email: david@essex.ac.uk;
lcarrea@essex.ac.uk).}}
%
%
%

\maketitle

\begin{abstract}
Geometric Polarimetry \cite{bebbington:GPI} has recently been introduced as a new analytical framework to express fundamental relationships in polarimetry, characterizing these in geometric terms which guarantees their invariance with respect to spatial reference frame and choice of basis. It was shown via a rigorous derivation from Maxwell's equations that there is a formal argument for representing elementary coherent states algebraically as spinors, and geometrically as generators of the Poincar\'{e} sphere.  While \cite{bebbington:GPI} only considered the characterization of field states, there is in remote sensing contexts a corresponding need also to characterize the polarization states of antennas.  This needs to be completely generic and not dependent on the detailed structure of the antenna.  This paper presents a derivation based on Schelkunov's reaction theorem \cite{schelkunoff:1939} which fulfils these requirements. The statement of the theorem is translated from its usual form to a tensor representation, and this is finally reduced to obtain the new spinor representation of the antenna in terms of its polarization spinor and its phase flag.
\end{abstract}


%
\IEEEpeerreviewmaketitle

\section{Introduction}

The response of an antenna to an incoming polarized plane electromagnetic wave may be posed in general as a scattering problem. It would require a detailed satisfaction of the boundary conditions over the surface of the antenna, and it would depend very much on the precise structure of the antenna.  Yet, it is accepted in the context of analytic field representations that for these purposes the antenna can be characterised simply as having an effective (complex) dipole equivalent, and that the voltage at the antenna terminals can be expressed by an inner product between the dipole vector and the field of the plane wave in the absence of the antenna \cite{mottI}.

In recent work \cite{bebbington:GP0}, \cite{bebbington:GPI} certain problems in the conventional representations of complex analytic electromagnetic waves were highlighted. In particular, it has been shown that the traditional interpretation of Jones vectors as subject to unitary SU(2) basis transformations leads to difficulties when inner products are involved because Euclidean inner products are not preserved under such transformations. In the past, various arguments were presented over the representations of waves propagating outward or inward, of which the consimilarity representation became widely accepted, despite the fact that in general media it is physically invalid, as shown in \cite{bebbington:GP0}. In \cite{bebbington:GPI} a new, rigorous analytical representation of complex analytical polarization states was presented within a framework that we have called Geometric Polarimetry. The aim of this framework is to provide not only a rigorous but also a logically and physically consistent representation of polarimetric waves in both arbitrary bases, and with arbitrary propagation directions.  The frame-invariant nature of the theory is underpinned by the fact that in this framework fields and representations of them can all be expressed geometrically, as can relations between them.  The most interesting and fundamental departure from the traditional picture to emerge from that work was that `elementary' polarizations states can be represented algebraically as spinors, and geometrically as the one of the two families of complex line generators of the Poincar\'{e} sphere, which now geometrically can be overlaid on the sphere of unit wave vectors in real space. Conjugate wave vectors that enter into calculations of the covariance matrix of a coherent or incoherent field are naturally represented by the second, complementary set of generators, while real Stokes vectors (for a coherent state) are geometrically constructed by the intersection of mutually conjugate pairs which occurs at a single real point on the sphere.
The advantage of using spinor algebra is that its notation naturally distinguishes the transformational properties of each of the four different types of spinor, which can be divided into whether unprimed or primed (transformations mutually complex conjugate) or covariant or contravariant (mutually inverse).  In deriving the spinor representation of a plane wave pure polarization state (appearing like a Jones vector) directly from a tensor representation of the Maxwell field tensor it emerged that its natural representation is as an unprimed covariant spinor.   In radar and other forms of remote sensing instrumentation it is necessary to consider antennas, and this is where the departure from traditional approaches is most stark.  It is natural to expect that the voltage response of an antenna in reception to an incoming field will be a basis invariant scalar.  One then can posit that the voltage equation takes the form of an invariant spinor inner product,
\begin{equation}\label{Vpsieta}
    V=\psi_A\eta^A,
\end{equation}
where $\psi_A$  is the covariant representation of the incoming field and  $\eta^A$ would be the contravariant representation of the antenna state.  Logically this is the only possibility for an invariant scalar result, but this formulation appears to run as contradictory to the accepted wisdom that the antenna height vector is `the same' as the field it transmits.  If antenna states are contravariant while field states are covariant, such a statement cannot be accepted as formally rigorous.   The apparent contradiction can be explained away formally in the following framework.  Firstly, we accept analytical Jones vectors as complex extensions of Cartesian vectors.  We also accept the effect in the far field of an antenna as that of an equivalent dipole.  The evaluation of the field as a function of the source involves the action of the dyadic Greens function \cite{bebbington:GP0}, which amongst other things operates on the contravariant dipole (extensive in length) to produce a covariant field (evaluated in volts/length). In Cartesian algebra, in which the Euclidean metric is implicit, such distinctions are normally not made explicitly, because the effect of Euclidean metric which is numerically equal to an identity matrix appears to be invisible.   In a basic representation of fields and antennas in which the real and imaginary parts are referenced to real coordinate basis vectors, there is no problem with numerical equality (up to scale) between the antenna height vector and the field it radiates.  However, once unitary basis changes are contemplated, the argument breaks down because the Euclidean metric is no longer preserved under unitary transformations which are not Euclidean isometries.

To avoid the confusions that have arisen in the application of basis changes, we introduced in \cite{bebbington:GPI} a clear distinction between Jones vectors, proper vectors, and spinors, with the latter being the objects truly transformable under unitary transformations, and are geometrically fundamentally different entities (generating lines of the sphere) as opposed to the Jones vectors which retain a true vectorial identity.  Whilst \cite{bebbington:GPI} provided a rigorous derivation of wave spinors from Maxwell's equations, and (\ref{Vpsieta}) appears not only plausible but necessary, a complementary rigorous derivation of the antenna spinor state has as yet not been furnished.  The remainder of this paper is devoted to just that end.

To motivate the development, we note that in reciprocal media, an antenna has equal response in both transmit and receive mode.  By extension, it becomes clear that the receive properties considered as at the antenna are determined by the far field transmit response, simply by two complex amplitudes.  This can be justified on physical grounds by appealing to Schelkunoff's reaction theorem \cite{schelkunoff:1939}.  Reaction is a physical quantity that ultimately depends on two independent sources in a linear physical system.   If one considers an antenna to be in reception of a wave, then it is possible to drive a current through its terminals, and the resulting effect throughout space and in the antenna is a linear superposition of what happens with the distant source alone, and for the exciting current alone.  Reaction however involves products of the two effects.  There is a reaction $R$ at the antenna terminals,
\begin{equation}\label{R}
    R=<V_r,I_t>
\end{equation}
between the received voltage $V_r$ from the wave generated by a distant source, and the drive current $I_t$ applied to the terminals.   There is also a reaction term due to products of the incident field and the field that would be generated due to the drive current alone, integrated over a closed surface.  Schelknoff's reaction theorem is a form of statement of reciprocity that states that the reaction integrated through a volume is equal to a surface integral enclosing that volume.  The term $R$ in (\ref{R}) can, in effect be shrunk to a volume enclosing only the terminals, and is then equivalent to the reaction between two far fields: one being the incident wave far from the near field of the antenna, and the other being the field due to the antenna being driven.  Assuming the antenna radiation pattern is known, one can then deduce the antenna scalar voltage by dividing by the (linear) dependence on the drive current or equivalently, one can consider a unit test current.

As far as we are aware, in the literature Schelkunoff's reaction theorem has only been stated using traditional vector calculus.  To adapt it for the purposes of a derivation within Geometric Polarimetry it is necessary to restate it in tensor form.  Having done that, it is possible to utilize the fact that any tensor expression can be rendered in a straightforward way in spinor form.  Finally, one has to reduce the antenna from a vector representation (complementary to the Jones vector) to a pure 2-spinor.  As may be expected, the device of reducing it by means of a reference spinor (which, according to its function we denote as a phase flag).   There are now two types of phase flag: one associated with a wave, the other with an antenna.  We show that a relationship must exist between these two in any local reference frame for the relation (\ref{Vpsieta}) to be valid.

%

\section{Voltage and orthogonality using spinors} \label{corres}

In \cite{bebbington:GP0} we  set out a new explanation for Graves' congruential transformation rule avoiding non-physical conjugation of the wave states. Central to the derivation was the requirement that antenna height polarization vectors must transform under basis transformation by the inverse of the transformation for the received field spinor. The conclusion was that we must consider wave polarization states as covariant spinors while 
antenna polarization states must be regarded as represented by contravariant spinors. For a wave $\psi_A$ and an antenna state $\eta^A$, the received voltage may be defined to be:
\begin{equation}\label{V}
    V=\psi_A\,\eta^A.
\end{equation}
It is important to note that the two objects $\psi_A$ and $\eta^A$ are defined with respect to the same spin-frame basis. This is in any case a natural condition since it is implicit that the antenna height is referred to the direction of the wave. The above relationship is the only kind that makes the result fully covariant, in other words independent of basis change. Empirically, one could obtain $\eta^A$ by differentiation, i.e.
\begin{equation}\label{}
    \eta^A=\frac{\partial\, V}{\partial\, \psi_A},
\end{equation}
which is entirely consistent in terms of the honouring of spinor character.


One of the compelling reasons  for choosing to develop Geometric Polarimetry using the language of spinors is that the algebra appears to work in precisely the way it should, something that will become increasingly evident as this work proceeds. It has at least been evident that the way polarimetric entities are annotated must be changed to make sense of the structure, and hence it has seemed preferable to adopt an already existing notation rather than invent one that has no prior literature.  There is, however one facet that is potentially confusing as a carry-over from conventional usage, which has to do with the concept of orthogonality. Since any inner product of the form 
\begin{equation}\label{}
    \psi_A\psi^A=\varepsilon_{AB}\psi^A\psi^B,
\end{equation}
with $\varepsilon_{AB}$ the usual skew metric spinor \cite{bebbington:GPI}, is identically zero, this expression does not in spinor form connote geometric orthogonality.  The use of identical labels for spinors related by the metric spinor in this way appears to sit uneasily with the concept of antennas and the wave states they respond to maximally being allocated under the IEEE convention as 'belonging' to the same state of polarization \cite{ieee83}.  To resolve this, we propose a minor terminological innovation, that when an expression involving pairs of spinors vanishes, we say that they are \emph{corresponding} states. Then, for example, a horizontally polarized antenna, \emph{corresponds} (or, for emphasis, is in \emph{null-correspondence}) with a vertical polarized wave (relative to the metric spinor). This concept may be generalised so that if there is a null return from a scatterer, the transmit and receive states may then be said to correspond with respect to the scatterer. This may initially seem contrived, but it is necessary because the prior conceptual framework does not accord with the actual mathematical framework that applies.  It is worth remembering that polarimetry already encompasses the idea of orthogonal states being antipodal on the Poincar\'{e} sphere.  The idea of correspondences is common in projective geometry, and it will be found helpful to use this in many instances, because significant geometric relationships are often most clearly expressed in this way.

\section{Derivation of the complex antenna height spinor} \label{derivation}

Given the controversies that have arisen over decades in the field of radar polarimetry, it seems important to go beyond setting out axioms and develop all our theory from first principles.  In \cite{bebbington:GPI} we established the first half of the premise by deriving the wave polarization spinor in terms of the tensor representation of the Maxwellian field. We now derive the antenna height spinor from fundamental principles. In this case it has to be recognised that antennas come in many variants so one must appeal to a general principle to establish the result for an arbitrary antenna. Schelkunoff's reaction theorem \cite{schelkunoff:1939}, \cite{schelkunoff:1943}, \cite{rumsey:PhysicalReview}, \cite{richmond} provides the necessary springboard for the analysis, which in turn appeals to the Lorentz reciprocity theorem \cite{lorentz}, \cite{carson}.  These are best known expressed in the traditional $3-$vector calculus, which we will in due course translate into tensor form.  Basically we consider an incident plane wave $\mathbf{E}_i$ which is supposed to induce an open circuit voltage $V$ across the terminals of a receiver antenna. We then suppose the receiving antenna to be driven by a test current, $I$, which would, in the absence of the initial field, result in a radiated test field $\mathbf{E}_t$. According to Schelkunoff's theorem \cite{schelkunoff:1939}, \cite{schelkunoff:1943}, \cite{rumsey:PhysicalReview}, \cite{richmond} the reaction $<V,I>$ between the fields is,
\begin{equation}\label{VI}
    <V,I>=\int\limits_S\!\!\!\!\int \left(\mathbf{E}_t\times \mathbf{H}_i- \mathbf{E}_i\times \mathbf{H}_t\right)\cdot \mathbf{n} \,dS
\end{equation}
where $\mathbf{H}_i$ and $\mathbf{H}_t$ are the associated magnetic fields, and $S$ is an arbitrary surface enclosing the antenna. In practice one would chose a spherical surface in the far field of the antenna. The reaction can be considered as the instantaneous power transferred to the source by the test current in the receive antenna, so the voltage can be derived if the right-hand side is known. It is important to note here that no conjugations are employed. Some care needs to be taken with the signal representation for these fields. Strictly one should consider real fields and the result is not time invariant but contains terms at the sum frequency of the two waves analogously to the real Poynting vector. It is possible to relax the condition on a real representation and use analytic signal representations by keeping strict track of the terms arising from the full positive and negative frequency expansion of the relevant sinusoids. If the underlying harmonic frequency is $\omega$ the result must include terms for harmonic frequencies $-2\omega$, $0$, $+2\omega$, giving three independent scalar equations encoding the result. It becomes clear that use of the positive analytical frequency in each term on the right hand side gives a meaningful result for the received voltage having the same analytic signature as the incident field when finally dividing out the analytical current signal.

\begin{figure}[h]
\centering
\includegraphics[scale=0.45]{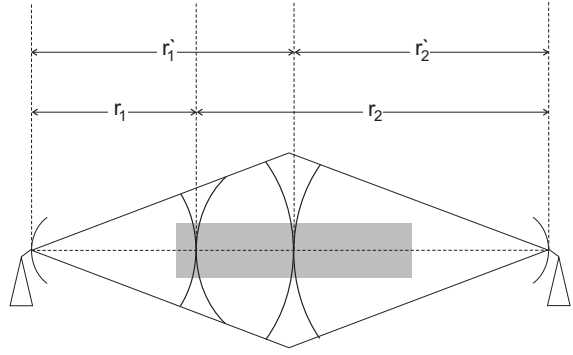}
\caption{For the reaction integral the sum of the phase factors on each stationary surface (in shaded region the far field of each antenna) are constant. So the reaction integral is range independent.}
\label{Reaction}
\end{figure}

\begin{figure}[h]
\centering
\includegraphics[scale=0.4]{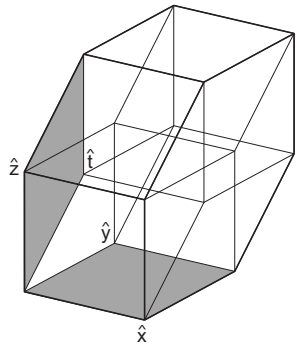}
\caption{Unit $4-$volume element of space-time spanned by unit $2-$planes $\hat{x}\wedge\hat{y}$ and $\hat{z}\wedge\hat{t}$.}
\label{Hypercube}
\end{figure}

Anticipating the result on the assumption that the contributions to the integral are only significant for angles where the incoming plane wave and test signal are almost in opposite directions, it is clear that the phasor product, $\mbox{e}^{j(\omega t-kz)}\,\mbox{e}^{j(\omega t+kz)}$ is essentially independent of the radius of the sphere centred on the antenna (see Fig. \ref{Reaction}). On the usual assumption that the vectorial factors in (\ref{VI}) vary much more slowly than the phases, we can take them out of the integral and integrate the angular dependencies. The phase difference between a plane wave and spherical wave is quadratic in the offset from the paraxial direction, and hence the angular surface phasor integral is closely approximated by a $2-$dimensional Fresnel integral, when expressed in polar coordinates. We then have that
\begin{equation}\label{}
    r^2\,\int\!\!\!\!\int \mbox{e}^{jkr(1-\cos\theta)}\,\sin\!\theta \,\,d\theta \, d\varphi=\frac{2\pi j}{k}r,
\end{equation}
for large enough $kr$. The apparent radial dependence on the right-hand side is simply the compensating factor for the $\frac{1}{r}$ dependence on the far-field amplitude of the test signal in the integrand, confirming the arbitrariness of the surface, $S$.
Turning now to the vector term, we require a spinorial representation to complete the derivation. The electromagnetic spinor was derived in \cite{bebbington:GPI} from the tensor representations of the magnetic field, so the first step is to convert the usual $3-$vector form of the reciprocity integrand to tensor form.  The Maxwell electromagnetic field tensor \cite{bebbington:GPI} is an example of a real bivector and it is given in covariant form by,
\begin{equation}\label{Fab}
    F_{ab}=\begin{pmatrix}
  \quad\!0 & \quad\!E_x & \quad\!E_y & \quad\!E_z \\
  -E_x & \quad\!0 & -H_z & \quad\!H_y \\
  -E_y & \quad\!H_z & \quad\!0 & -H_x \\
  -E_z & -H_y & \quad\!H_x & \quad\!0 \\
\end{pmatrix}.
\end{equation}
Customarily, when writing $F_{ab}$ in this form the units are assumed implicitly to be homogenized by including a factor equal the impedance of free space in the magnetic field. Denoting the incident field by $F$ and the test field by $F'$, it is found that the vector part of the integrand may be expressed as
\begin{eqnarray}\label{integrand}
  && \left(\mathbf{E}_t\times \mathbf{H}_i- \mathbf{E}_i\times \mathbf{H}_t\right)\cdot \mathbf{n} \,dS = \\
  && = \left(F_{ab}\,F'^b_{\:\:\:\:c}-F'_{ab}\,F^b_{\:\:\:c}\right)\,\varepsilon^{ac}_{\:\:\:\:\:lm}\,\,dx^l\,dx^m , \nonumber
\end{eqnarray}
where $\varepsilon^{ac}_{\:\:\:\:\:lm}$ is the usual antisymmetric Levi-Civita symbol \cite{penrose} which is
\begin{equation}\label{}
    \varepsilon^{ac}_{\:\:\:\:\:lm}=\left\{\begin{array}{l}
                                            +1 \mbox{ for even permutations of 0123} \\
                                            \quad \!0 \mbox{ if any index is repeated} \\
                                            -1 \mbox{ for odd permutations of 0123}
                                          \end{array}\right.
\end{equation}
and index raising (lowering) are performed by the Minkowski metric tensor $g^{ab}$ ($\,g_{ab}$), where,
\begin{equation}\label{}
    g^{ab}=g_{ab}=\begin{pmatrix}
      1 & \quad\!0 & \quad\!0 & \quad\!0 \\
      0 & -1 & \quad\!0 & \quad\!0 \\
      0 & \quad\!0 & -1 & \quad\!0 \\
      0 & \quad\!0 & \quad\!0 & -1 \\
    \end{pmatrix}.
\end{equation}
From here we may specialise the form of the Maxwell field tensor to the case of harmonic waves, such that it may be related to the vector $4-$potential $\Phi_a$ and the covariant $4-$vector of the wave, $k_a=\left( \frac{\omega}{c},k_x,k_y,k_z\right)$ \cite{bebbington:GPI},
\begin{equation}\label{}
    F_{ab}=j(k_a\Phi_b-k_b\Phi_a).
\end{equation}
We now substitute this form in (\ref{integrand}) to express the integral in terms of the vector potentials, bearing in mind that the vector parts of the two wave vectors are antipodal. The term in brackets in (\ref{VI}) becomes
\begin{equation}\label{}
    k'_b\,k^b\left(\Phi'_a\Phi_c-\Phi_a\Phi'_c\right)+\Phi'_b\,\Phi^b\left(k'_ak_c-k_ak'_c\right).
\end{equation}
When integrated over the area element, the first term vanishes because it is purely transverse, while if the directed area is in the $xy$ plane, the factor in the second term is a bivector in the $tz$ plane orthogonal to it, so these together create a non-vanishing $4-$volume element shown in Fig. \ref{Hypercube}. We therefore have that the reaction reduces to the form,
\begin{equation}\label{VIPhi}
    <V,I> \,\propto \Phi_b\,\Phi'^b = \Phi'_b\,\Phi^b.
\end{equation}
The symmetry of this relation reflects the fact that the reaction of the driving current at the source antenna due to the voltage induced by the incident field must be the same as that of the voltage induced in an antenna that is the source of the incident field by the test field. The required derivation is almost complete.  In \cite{bebbington:GPI} we showed how the wave state could be derived as a single index spinor by reducing the mixed index spinor for the potential, using the phase flag spinor,
\begin{equation}\label{psi}
    \psi_A=\Phi_{AA'}\bar{\theta}^{A'}
\end{equation}
which is a form of linear projection. The potential for the test source is evidently linked to the distribution of current sources in the antenna, and formally obtained using the dyadic Green's function \cite{bebbington:GP0}. The dyadic Green's function is a product of the scalar Green's function and a vector part that projects the elementary current direction into the transverse plane. In spinor form the same function is realised using combinations of spinors based on the spin frame of the wave propagation vector. Selecting a spin-frame basis for the standard $z-$propagating wave, as $\{o^A,\iota^A\}$, it can be verified that the operator,
\begin{equation}\label{PI}
    \Pi^{AA'}_{\;\;\;\;\;BB'}=-\,o^A\iota^{A'}\iota_Bo_{B'}-\iota^A o^{A'}o_B\iota_{B'}
\end{equation}
acts to nullify $o^Ao^{A'}=l^a$ and $\iota^A\iota^{A'}=n^a$, and is idempotent with respect to $o^A\iota^{A'}=m^a$ and $\iota^Ao^{A'}=\bar{m}^a$, where $l^a=t^a+z^a$, $n^a=t^a-z^a$, $m^a=x^a+j\,y^a$ and $\bar{m}^a=x^a-j\,y^a$ are the standard null tetrad coordinates expressed in terms of the usual orthogonal basis vectors of spacetime \cite{bebbington:GP0}.
The operator (\ref{PI}) therefore projects out the timelike and longitudinal components relative to the wave when the wave is aligned with the spin frame ($\{oA,\iota^A\}$ for $z-$propagating waves), and as required leaves the transverse components intact.  We may therefore express formally,
\begin{equation}\label{PhiAAp}
    \Phi^{AA'}=\frac{Z_0 I}{4\pi r}\,h^{AA'}=\int\!\!\!\!\!\!\!\!\int\limits_{Source}\!\!\!\!\!\!\!\!\int \frac{\mbox{e}^{j\,\mathbf{k}\cdot(\mathbf{r}-\mathbf{r}')}}{4\pi|\mathbf{r}-\mathbf{r}'|}\:\Pi^{AA'BB'}J_{BB'}\,dV
\end{equation}
as a working definition of the complex antenna height vector, which is transverse in the spin frame of the wave traveling from source to far field point of measurement. In the standard frame we arrive at a form for the test potential,
\begin{equation}\label{}
    \Phi'^{AA'}=\begin{pmatrix}
                  0 & \Phi'_l \\
                  \Phi'_r & 0 \\
                \end{pmatrix}
\end{equation}
with subscripts $l$ and $r$ referring to left and right hand circular polarization components. In unreduced form, this contracts with the potential for the field of the counterpropagating wave,
\begin{equation}\label{}
      \Phi_{AA'}=\begin{pmatrix}
                  0 & \Phi_l \\
                  \Phi_r & 0 \\
                \end{pmatrix}
\end{equation}
to obtain $<V,I>$ in (\ref{VIPhi}).
The left and right hand components end up in the same off-diagonal positions in each case because the wave directions for the two fields are in opposing directions; the interpretation of $\Phi_{01'}$ and $\Phi_{10'}$ as right or left handed polarizations is sensitive to the direction when we use the same spin-frame to represent both waves. Index raising involves applying the skew metric spinors $\varepsilon^{AB}$ and $\varepsilon^{A'B'}$. The fact that all the calculations here have been done with respect to a single frame of reference should be carefully noted. No tricks of wave reversal transformation have been involved.

The final step of our derivation is to find the analogue of (\ref{psi}), which reduces the antenna height vector in (\ref{PhiAAp}) to a spinor.  Assume we have a covariant phase flag, $\bar{\tilde{\theta}}_{A'}$  for the reduction of $h^{AA'}$, then (\ref{V}) must take the form,
\begin{eqnarray}\label{Vtheta}
  V &=& tr\left[\begin{pmatrix}
                0 & \psi_0\,\bar{\theta}^{1'} \\
                \psi_1\,\bar{\theta}^{0'} & 0 \\
              \end{pmatrix}\begin{pmatrix}
                0 & \eta^0\,\bar{\tilde{\theta}}_{1'} \\
                \eta^1\,\bar{\tilde{\theta}}_{0'} & 0 \\
              \end{pmatrix}^T\right]= \nonumber \\
   &=& \psi_0\,\eta^0(\bar{\theta}^{1'}\bar{\tilde{\theta}}_{1'}) + \psi_1\,\eta^1(\bar{\theta}^{0'}\bar{\tilde{\theta}}_{0'}  ),
\end{eqnarray}
where $tr$ expresses the trace of a matrix. To make this an expression that transforms covariantly under basis transformation there must be a fixed geometric relationship between the two phase flags such that the coefficients in the bracketed factors on the right-hand side of (\ref{Vtheta}) are equal. This condition should not be surprising because the phase flags refer to waves with counter-propagating wave vectors. It was already clear in \cite{bebbington:GPI} that there is a constraint on the phase flag that its vector should lie in the equatorial plane of the spin frame defined by the wave. The relation (\ref{Vtheta}) shows that a still tighter constraint should be made to ensure consistency. We require that both coefficients in (\ref{Vtheta}) should be equal. This condition is respected if firstly, the phase flag is fixed in the form,
\begin{equation}\label{}
    \bar{\theta}^{A'}=\bar{\kappa}^{A'}+\bar{\lambda}^{A'}
\end{equation}
where $\kappa^A\bar{\kappa}^{A'}$ is the normalised wave vector, and $\{\kappa^A,\lambda^A\}$ is the spin frame for the wave, and secondly that the spin frame for the test wave be fixed as $\{\lambda^A,-\kappa^A\}$. This simply fixes the relative rotations and sensibly ensures that the circularly polarized waves of either direction are deemed to have zero phase when the field is aligned with the same plane. As a result we have that, in the standard spin frame,
\begin{eqnarray}
  \kappa^A &=& \begin{pmatrix}
               1 \\
               0 \\
             \end{pmatrix}, \quad  \lambda^A=\begin{pmatrix}
               0 \\
               1 \\
             \end{pmatrix}, \\
  \theta^A &=& \kappa^A+\lambda^A=\begin{pmatrix}
               1 \\
               1 \\
             \end{pmatrix}, \quad \bar{\theta}^{A'} =\begin{pmatrix}
               1 \\
               1 \\
             \end{pmatrix} \nonumber
\end{eqnarray}
while
\begin{eqnarray}
  \lambda^A &=& \begin{pmatrix}
               0 \\
               1 \\
             \end{pmatrix}, \quad  -\kappa^A=\begin{pmatrix}
               -1 \\
               \quad \! 0 \\
             \end{pmatrix}, \\
  \tilde{\theta}^A &=& \lambda^A+(-\kappa^A)=\begin{pmatrix}
               -1 \\
               \quad\! 1 \\
             \end{pmatrix}, \; \bar{\tilde{\theta}}^{A'} =\begin{pmatrix}
               -1 \\
               \quad \!1 \\
             \end{pmatrix}, \; \bar{\tilde{\theta}}_{A'} =\begin{pmatrix}
               1 \\
               1 \\
             \end{pmatrix}. \nonumber
\end{eqnarray}
Hence the required condition $(\bar{\theta}^{1'}\bar{\tilde{\theta}}_{1'})=(\bar{\theta}^{0'}\bar{\tilde{\theta}}_{0'}  )$ is fulfilled.

To conclude the discussion of this section, it is entirely expected that there should be an inner product between complex antenna  states and wave states that is both basis-invariant and geometrically invariant, because voltage measured at the receive antenna is invariant.  The problem in polarimetry has always been about how to do this consistently using a unified formalism.  Though a somewhat elaborate technical approach, use of the reaction principle has provided a means to show the formalism can indeed express this  quite generally regardless of the details of the source or receive antenna.  We have also shown that there is a consistent way to define phase flags that reduce vector fields to the complex two-component (spinor) states that are traditional to polarimetric algebra.

\section{Conclusions}

The translation of the statement of Schelkunoff's reaction theorem to 4-tensor form proves to be quite straightforward.   Given the usual appeal to stationary phase integration common to many areas of scattering theory, including the derivation of the extinction theorem, the arbitrariness of the chosen surface of integration justifies the approach.  It is seen that the reaction depends only on an inner product of the  transverse components of the vector potentials generated by the two sources, and when phase flags for these are introduced, the received voltage takes the expected form as the inner produce of covariant and contravariant spinors representing respectively the received field and the receiving antenna.

The form of the phase flags is partly determined and partly a matter of convention.  It is only strictly necessary for the two components to have equal weight when expressed in a frame aligned with the direction of propagation.  This is to say that the components of the wave analysed into each handedness of polarization are weighted equally in their contribution to the resulting spinor.  Beyond this requirement the actual phases of the components of the phase flag are a matter of convention.   Physically and overall scalar phase factor is what physicists refer to as a U(1) gauge symmetry, and relates to what one considers as phase zero.  On the other hand a differential phase has the effect of retarding one of the circular polarizations while advancing the other.   Given the obvious (circular) statement that the circularity of the polarization ellipse for pure circular polarization means that there is no physically distinguished angle that can be identified to mark zero phase, it can be seen that the adoption of a phase flag convention is really equivalent to determining one direction in the transverse plane of the wave where the phase of the instantaneous circularly polarized field is zero.  Our choice of convention equates to stating that for radiation along the $z$ axis both left and right components are considered at time zero to be real in the ($xy$) aperture plane where they cross the $xz$ plane.  This construction can be expressed in a universal manner by expressing the phase flag in terms of the basis spinors of the spin frame characterizing the frame in which the antenna measures the field.   There are subtleties in this statement, long known since the work of Ludwig \cite{ludwig} when one considers detailed evaluation of antenna patterns extended over finite angles, or indeed when an antenna is considered to be steerable.  For then, one has to consider not only how the pointing direction of the antenna is moved, but also the effects of any rotation of the transverse coordinate system. The effect of arriving at the same point by a different routes in configuration space is one of phase holonomy, well known in optics as an example Berry's phase \cite{berry96}, \cite{berry87}. The advantage of the phase flag is that by applying the unitary representations of a sequence of rotations the induced Berry phase can readily be computed.




\section*{Acknowledgment}
The work on Geometric Polarization was supported by the U.S.
Office of Naval Research and the European Union (Marie Curie
Research Training Network RTN AMPER, Contract number
HPRN-CT-2002-00205). We also owe a debt of gratitude in particular
to Prof. Wolfgang Boerner and the late Dr. Ernst Lueneburg, who
were prime movers in stimulating the debate on the foundations of
polarimetry, regardless of where that might lead.


\bibliographystyle{IEEEtran}
\bibliography{IEEEabrv,bibliografia}

\newpage

%
%
%

\end{document}